\documentclass[a4paper,11pt]{JHEP3}
\usepackage{amsmath}
\usepackage{epsf, amssymb}
\usepackage{epsfig}

\title{\Large{Type II Small Stringy Black Holes, Probe Branes and Higher Derivative Interactions}}
\author{Ling-Yan Hung and Linda I. Uruchurtu \\ \\
        Department of Applied Mathematics and Theoretical Physics \\
        University of Cambridge \\
        Wilberforce Road, Cambridge CB3 0WA, United Kingdom.\\
        E-mail: \email{lyh20@damtp.cam.ac.uk}\\
        \hspace{13.mm}\email{liu20@damtp.cam.ac.uk}}

\abstract{The near horizon geometry of a fundamental string wrapped
around an $S^{1}$ reduced to four dimensions
is expected to be $AdS_{2}\times S^{2}$.
A probe string analysis suggests a no-force condition indicating
supersymmetry, which coincides with the condition that the $AdS_{2}$ is embedded in $AdS_{3}$.
We therefore consider the bulk string theory in terms of a WZW model on $AdS_{3}$
following recent proposals by Dabholkar et. al and Giveon et. al. We find that conformal
symmetry of the model naturally leads to the no-force
constraints obtained from the probes. %, which surprisingly does not receive further $\alpha'$ corrections.
Moreover, we are able to extract the values of the moduli that account
for the value of the microscopic entropy.
We also investigate higher derivative  corrections of the form
$\alpha'^3\mathcal{R}^{4}+\mathrm{flux\hspace{2.mm}terms}$ to the
horizon, in the context of type IIB supergravity. Imposing the no-force condition from the probe
analysis leads to a striking simplification of the equations of motion at this order
in $\alpha'$. However, we argue that the value of the entropy can only be determined by
considering all orders in $\alpha'$.
}

\preprint{DAMTP-2007-118}


\begin{document}

\newpage

%%%%%%%%%%%%%%%%%%%%%%%%%%%%%%%%%%%%%%%%%%%%%%%%%%%%%%%%%%%%%%%%%%%%%%%%%%%%%%%%%%%%%%%%%%%%%%%%%%%%%%%%%%%%%%
\section{Introduction}
Ideas concerning duality in the context of
supergravity and string theory, have modified our understanding of
classical and quantum aspects of black holes. In particular string theory has been successful in
providing a correct microscopic description of the entropy of
various classes of black holes \cite{Strominger:1996sh}, in accord with the Bekenstein-Hawking formula.

Several classes of multi-charged black holes in
string theory have illustrated a match between microscopic and macroscopic
entropies \cite{Strominger:1996sh,Cvetic:1995kv,Cvetic:1995bj}.
However there has been a puzzle regarding extremal black holes with two
or less charges, which are usually referred to as small black holes.
These have vanishing horizon area, and thus zero Bekenstein-Hawking
entropy. In order to solve this apparent paradox, it is argued that
the vanishing horizon can be traced back to the fact that the
curvatures near the core of these black holes are very large, so
that a horizon of the string scale could only be seen after including
the higher derivative terms so far neglected \cite{Dabholkar:2004dq}. This was shown
explicitly in \cite{Sen:2005kj} for the heterotic winding string
with momentum compactified to four or five dimensions. It was a remarkable coincidence that in this case
the full entropy obtained from microscopic
counting was reproduced by simply adding the Gauss-Bonnet term
\cite{Sen:1995in,Sen:2005iz} to the classical supergravity action \footnote{For $D>5$ one needs
to consider higher extended Gauss-Bonnet densities, which are of higher order in $\alpha'$. This has been
discussed in \cite{Cvitan:2007en}}.
The assumption behind this is that after
supersymmetrizing this term (which has not been done), the related
flux terms will not alter the conclusion. Higher order $\alpha'$
effects are also not expected to change the result although this has
not been proved. It is then our main aim to look for
the analogous statements for the type II fundamental string black
holes.

In this paper we will study features of the two-charge black hole obtained
by wrapping a type II fundamental string $m$ times around a circle
with momentum $p=n/R$, where $R$ is its radius. The
system is then compactified on $T^{5}$
to obtain a black hole in four dimensions. Such a configuration is supersymmetric
and has been shown to have a non-zero entropy given by
$S=2\pi\sqrt{2nm}$.  However, it is well known that the classical solution
corresponding to such a black hole has vanishing horizon and therefore,
zero Bekenstein-Hawking entropy. Hence, it is expected that higher $\alpha'$
corrections will radically alter the horizon geometry.

Since we do not have an explicit solution for a black hole with a stretched
horizon in type II supergravity, we will follow Sen and assume that
any extremal black hole must have
a near-horizon geometry with an $AdS_{2}$ isometry. Clearly supersymmetry
will provide further constraints on the geometry, but so far the $\alpha'$
corrected Killing spinor equations remain unknown. However, it is expected 
that appropiate probe strings and D-branes must experience no force.
Since the background itself is supposed to be generated by a string wrapping a compact
circle, the most natural probe one can use, is a string
parallel to the background string. It is then our strategy to test the background
with strings in order to determine a BPS condition relating the moduli (radii of the
horizon metric and fluxes). Following the known giant
graviton literature \cite{McGreevy:2000cw,Grisaru:2000zn,Hashimoto:2000zp},
we also considered dual probes which wind the sphere transverse to the background string,
and tried to look for BPS-like conditions. We will find that these
conditions restrict the black hole moduli in such a way, that the
$AdS_{2}\times S^{1}$ is enhanced to $AdS_{3}$. 

The connection between $AdS_{3}$ and $AdS_{2}$
is familiar from the literature on five dimensional heterotic black holes (see \cite{Strominger:1998yg, Kraus:2006wn}).
However, our probe analysis shows that the symmetry enhancement is also realised in the context
of type II black holes, and furthermore motivates the possibility of looking at the fundamental string worldsheet
theory being the holographic dual of this $AdS_{3}$ black hole geometry. This idea has
been explored recently in the literature \cite{Giveon:2006pr,Dabholkar:2007gp,Lapan:2007jx, Kraus:2007vu}. In
particular in \cite{Giveon:2006pr,Dabholkar:2007gp} the bulk geometry is
given by some $SL(2,\mathcal{R})$ WZW model at level 2 which reproduces the symmetries of the fundamental
string worldsheet CFT. Given that in this proposal the bulk
geometry is an exact CFT, one can read off the values of the moduli
from the WZW action, and these values should be correct to all orders in $\alpha'$. 
We will see that indeed, the
WZW action implies the no-force conditions we found from the probe analysis, and that the
values of the moduli reproduce the full entropy of the black hole, when they are
substituted into the entropy function. The fact that one gets full agreement with microscopic counting,
supports Sen's scaling argument for computing higher derivative corrections to the entropy. Furthermore, 
the no-force conditions are recovered from a particular WZW action describing strings in $AdS_{3}$
which supports the $AdS_{3}/CFT_{2}$ construction for the type II fundamental string.

We will then explicitly consider the effect of higher derivative corrections
relative to the Einstein-Hilbert action, and investigate the effects of field
redefinitions. In type II string theory, corrections to the supergravity action
start at order $\alpha^{'3}$, rather than $\alpha'$ as
in the heterotic case. Also, unlike the heterotic case where the
coefficient of the Gauss-Bonnet term is unity (with an appropriate
normalization), the coefficient of the type II $R^4$ term involves
an irrational $\zeta(3)$ factor. As a result, it is expected
that the full entropy will not be reproduced in this case.  In
addition there are field redefinition ambiguities
\cite{Tseytlin:1986zz} at this order which could enter into the
determination of the entropy. These arise from the fact that string
scattering amplitudes can only determine the coefficients of the
terms in the action that are not proportional to the lowest order
equations of motion. As a result, the undetermined coefficients are
ambiguous, and need to be fixed by other arguments, such as absence
of ghosts \cite{Zwiebach:1985uq} or ``off-shell'' supersymmetry \cite{Mohaupt:2000mj}.
These issues also enter in any kind of type II black hole setup. For
instance, in \cite{Ghodsi:2006cd} a particular
class of field redefinitions was explored while considering corrections to the
D1-D5p black hole, which has three charges and non-zero size. It was
argued that the $\alpha'^3$ corrections could be treated perturbatively,
so that after linearizing the attractor equations, the entropy was free of any ambiguity.
In \cite{Sinha:2006yy},  a single charge black hole was
studied. In this case, the black hole has zero size and the $\alpha'^3$ corrections cannot be
treated perturbatively. Because of this, the authors chose a particular
field redefinition in which the terms proportional to the Ricci
tensor, were removed, and showed that the horizon was stretched.

We will investigate how the field redefinition parameters affect the
entropy at order ${\alpha'}^{3}$ and explore the special role
played by the no-force conditions in distinguishing supersymmetric
solutions from non-supersymmetric ones.  We focus our
attention on two cases. We first truncate the action and consider
corrections coming only from the  gravitational sector in the dimensionally
reduced theory. Then we take into account the flux terms by using the
conjecture in \cite{Gross:1986mw,Kehagias:1997cq}. We will find that
restricting the corrections to the gravitational sector is in general
inconsistent with the no-force conditions obtained from the probe analysis.
However, once one includes contributions from the gauge
fields, and imposes the no-force constraints, the attractor equations simplify
drastically. This occurs independent of field redefinitions and allows one to solve for
the moduli for each choice of field redefinition parameters. Yet, the entropy
will depend on them, suggesting  that for the type II small black
hole, one needs to consider all $\alpha'$ corrections, in contrast to the heterotic
case. This reinforces the need to treat the bulk geometry using the
WZW exact CFT.

The plan of this paper is as follows. In section 2, we review the
Kaluza-Klein compactification of type IIB supergravity, from ten to
four dimensions, and we focus on the case of a winding string
coupled to the Neveu-Schwarz field. In section 3, we briefly
introduce Sen's entropy formalism for determining the entropy from
the knowledge of the effective action, and apply it to our case of
study. Section 4 presents a family of probe string and probe D2-brane
solutions in the $AdS$ background and the derivation of the
no-force conditions. In section 5 we
discuss the values of the moduli in the WZW model proposed in
\cite{Giveon:2006pr,Dabholkar:2007gp} and the resulting entropy from Sen's entropy
formalism. Finally section 6 is devoted to the computation and
analysis of the higher derivative corrections to the entropy.
Conclusions are then presented in section 7.
%%%%%%%%%%%%%%%%%%%%%%%%%%%%%%%%%%%%%%%%%%%%%%%%%%%%%%%%%%%%%%%%%%%%%%%%%%%%%%%%%%%%%%%%%%%%%%%%%%%%%%%
\section{Type IIB Action in Four Dimensions}
%%%%%%%%%%%%%%%%%%%%%%%%%%%%%%%%%%%%%%%%%%%%%%%%%%%%%%%%%%%%%%%%%%%%%%%%%%%%%%%%%%%%%%%%%%%%%%%%%%%%%%%
The known supergravity solution of a BPS fundamental string winding
a compact circle with some momentum $p$ around the circle has, like
many other extremal black brane solutions, vanishing horizon area
\cite{Dabholkar:1989jt}. However, albeit in a different regime in
coupling constants, it is possible to do a microscopic counting of
string states of the system and get a finite
result for given winding number $w$ and momentum $p$ according to
the Cardy's formula. It is suspected that by including $g_s$ and $\alpha'$ corrections,
to the supergravity Lagrangian, the horizon of the black hole becomes ``stretched'',
so the area and subsequently the Bekenstein-Hawking entropy, is non-vanishing
\cite{LopesCardoso:1998wt,LopesCardoso:1999xn,LopesCardoso:1999ur,Dabholkar:2004dq, Castro:2007sd, Castro:2007hc}.
A standard way for obtaining the corrections to the black hole
entropy from higher correcting terms in the Lagrangian was
devised in \cite{Sen:2005kj}. The basic assumption that underlies
the arguments is that the geometry of an extremal black hole in $D$
dimensions should be $AdS_2 \times S^{D-2}$ in the near horizon
limit. To obtain a black hole for our system, namely a type IIB
fundamental string winding around a compact circle with momentum, we
can consider compactifying it on a five-torus and a circle i.e. $T^5
\times S^1$ to obtain a black hole in four-dimensions \footnote{For
heterotic fundamental string it is known that when considering four
derivative corrections to the action, one obtains a black hole with
a near horizon geometry of $AdS_{2}\times S^{2}$
\cite{Dabholkar:2004yr,Dabholkar:2004dq}. For the type IIB case, no
explicit solutions have been obtained. }. In the near horizon limit
we should expect the metric to take the form \cite{Sen:2005iz}
\begin{equation}\label{metric1}
ds^2 = a\left(-r^2dt^2 + \frac{dr^2}{r^2}\right)+ b d\Omega_2^2
\end{equation}
where $a$ and $b$ to be determined by extremalising the action
after substituting in the corresponding metric.
The system carries the electric charges of NS two form $B_{\mu\nu}$ and
the Kaluza-Klein potential $A_\mu$ and we assume that at the horizon
the field strengths of these potentials take constant values $e_1$ and
$e_2$ respectively. The scalar fields including the (four dimensional)
dilaton and the field parametrising the radius of the compact circle
are all assumed to take constant values at the horizon.
\begin{equation}
e^{-2\phi} = u_s,
\qquad
e^{\frac{\psi}{2}} = u_T.
\end{equation}
In order to evaluate $f$ defined by \cite{Sen:2005kj}
\begin{equation}
f(\vec{u},\vec{v},\vec{e}) = \int_H d^Hx \sqrt{-\det g} L,
\label{ffunction}
\end{equation}
we need to reduce the ten-dimensional type IIB action to four
dimensions. We are reducing the theory on $T^5 \times S^1$ and there
are simple relationships between the higher dimensional fields and
their counterparts after reductions. These relations\footnote{These
relations assume (-+++) signature of the metric. Signs in front of
$e^{\psi}$ have to be inverted had we adopted the alternative
signature. There are also variations in the definition of the
dimensionally reduced dilaton too. In \cite{Sen:2005iz} for example
the dilaton is shifted whereas in \cite{Kiem:1997qj} it stays
unchanged. We will adopt the convention in \cite{Sen:2005iz}.} are given
by \cite{Kiem:1997qj}
\begin{equation}
\hat{G}_{\mu\nu} =
\left(\begin{array}{ccc}
g_{ab} + e^{\frac{\psi}{2}}A^{2}_{a}A^{2}_{b} & & g_{ty}   \\
          & R^2\delta_{mn} &\\
g_{yt}    & & e^{\psi}
\end{array} \right) ,
\end{equation}
where
\begin{eqnarray}
\hat{G}_{tt} & = & g_{tt} +  e^{\psi}(e_2r)^2  \nonumber \\
\hat{G}_{ty}       & = &  e^{\psi}(-e_2 r)      \nonumber     \\
\hat{B}_{ty} & = & -e_1r    \nonumber \\
\hat{\phi}   & = & \phi + \frac{1}{4}\psi
\label{Bfield}
\end{eqnarray}
where the hat denotes the ten-dimensional fields. The lower case latin
letters $a,b \in
\{0,1,2,3\}$ and $m, n$ runs from 4 to 8. $y$ denotes the
direction along the compact circle. $R$ is the radius of the flat torus and
is an arbitrary constant which could be absorbed in the definition of
the four dimensional dilaton and conveniently set to one. The Kaluza-Klein
potentials are given by $A_a = (-e_2r, 0,0,0)$ near the horizon because we have
assumed the field strength to be
\begin{equation}
F^{(2)}_{tr} = \partial_t A^{(2)}_r -  \partial_r A^{(2)}_t = e_2.
\end{equation}
The solution is static with no time dependence so we can set
$A_r=0$. Similarly we have
\begin{equation}
F^{1}_{tr} = \hat{H}_{try} = e_1 = (d\hat{B})_{try}.
\end{equation}
thus giving (\ref{Bfield}) above. Now we are ready to obtain the four
dimensional reduced Lagrangian from the ten dimensional one. This is
done by substituting the ten dimensional fields in terms of the four
dimensional ones using (\ref{Bfield}).
For example, the Ricci scalar in ten dimensions is reduced to
\begin{equation}
R^{(10)} = R^{(4)} - \frac{1}{4} e^{\psi}F_{(2)}^2.
\end{equation}
The lowest order effective type IIB action in four dimensions in the
NS sector is  given by
\begin{equation}
S_{IIB(4)} = \frac{2\pi V_5}{16\pi G_{10}}\int d^4x \sqrt{-g_{4}}
e^{-2\phi}\left[R^{(4)} - \frac{1}{4} e^{\psi}F^2_{(2)}  - \frac{1}{4} e^{-\psi}F^2_{(1)}\right]
\end{equation}
where we have omitted all the terms involving covariant derivatives of
the scalars and all components of $\hat{H}_3$ except $\hat{H}_{try}=
F^{(1)}_{tr}$, which are assumed to be zero at the horizon. We can
repeat the same tricks with the $R^4$ corrections.
%%%%%%%%%%%%%%%%%%%%%%%%%%%%%%%%%%%%%%%%%%%%%%%%%%%%%%%%%%%%%%%%%%%%%%%%%%%%%%%%%%%%%%%%%%%%%%%%%%%%%%%%%%
\section{Entropy Function Formalism}
%%%%%%%%%%%%%%%%%%%%%%%%%%%%%%%%%%%%%%%%%%%%%%%%%%%%%%%%%%%%%%%%%%%%%%%%%%%%%%%%%%%%%%%%%%%%%%%%%%%%%%%%%%
Here we review Sen's entropy formalism for the case of a fundamental
string winding a circle, on type IIB supergravity. We are unable to
work in ten dimensions, as there are no known regular solutions for
this system \cite{Dabholkar:1989jt,Peet:1995pe}. However, one can
follow Sen and assume that upon compactification, the fundamental
string will form an extremal black hole in four dimensions, with a
near-horizon geometry of $AdS_{2}\times S^{2}$. This metric can then
be lifted to ten dimensions, on $AdS_{2}\times S^{2}\times T^{6}$,
following the formulae in the previous section.

We consider an extremal black hole solution with near horizon geometry
\begin{eqnarray}
ds^2&=&a\left(-r^2dt^2+\frac{dr^2}{r^2}\right)
+b(d\theta^2+\sin^2\theta d\phi^2) \nonumber \\
H_{rty}&=&e_{1}, \hspace{10.mm} F_{rt}=e_{2} \nonumber \\
S&=&e^{-2\phi}, \hspace{8.mm} T=e^{\psi/2}
\end{eqnarray}
Using this background geometry, we compute the function defined by (\ref{ffunction}).
Since we are in the string frame, there
is an overall factor of $S$, and by re-scaling the fields  $f$ can be written as
\begin{eqnarray}
f&=&Sg(a,b,c,d)
\nonumber \\
c&=&e_{2}T \hspace{10.mm} d=\frac{e_{1} }{T}
\end{eqnarray}
The usual procedure is to extremize $f$ with respect to the moduli and
to determine the values of them. One then gets the set of equations
\begin{eqnarray}
\frac{\partial f}{\partial S}&=&0 \hspace{5.mm}\leftrightarrow\hspace{5.mm} g(a,b,c,d)=0
\nonumber \\
\frac{\partial f}{\partial a}&=&0 \hspace{5.mm}\leftrightarrow\hspace{5.mm} \frac{\partial g}{\partial a}
=0
\nonumber \\
\frac{\partial f}{\partial b}&=&0 \hspace{5.mm}\leftrightarrow\hspace{5.mm} \frac{\partial g}{\partial b}=0
\nonumber \\
\frac{\partial f}{\partial T}&=&0 \hspace{5.mm}\leftrightarrow\hspace{5.mm} c\frac{\partial g}{\partial c}
=d\frac{\partial g}{\partial d}
\label{attractor}
\end{eqnarray}
These equations can then be solved for $a,b,c,d$. The next step
is to compute the Legendre transform of this function, with respect to
the variables $(e_{1},e_{2})$. This is
\begin{equation}
F(a,b,q)\equiv 2\pi(e_{i}q_{i}-f(a,b,e_{i})), \hspace{10.mm} q_{i}
=\frac{\partial f}{\partial e_{i}}
\end{equation}
where
\begin{eqnarray}
q_{1}&=&\frac{\partial f}{\partial e_{1}} \hspace{10.mm}  q_{1}
=\frac{S}{T}\frac{\partial g}{\partial d}
\nonumber \\
q_{2}&=&\frac{\partial f}{\partial e_{2}} \hspace{10.mm}  q_{2}
=S T\frac{\partial g}{\partial c}
\end{eqnarray}
Using the equations (\ref{attractor}) and the explicit forms of the
$q_{i}$'s above, it is possible to express $S$ and $T$ as
\begin{equation}
S=\frac{\sqrt{q_{1}q_{2}}}{\sqrt{\frac{\partial g}{\partial c}\frac{\partial g}{\partial d}}}
\qquad
T=\sqrt{\frac{q_{2}}{q_{1}}\frac{\frac{\partial g}{\partial d}}{\frac{\partial g}{\partial c}}}
=\sqrt{\frac{q_{2}}{q_{1}}\frac{c}{d}}
\end{equation}
Evaluation of the function $F(a,b,q_{i})$ yields the value of the entropy. In this case
\begin{equation}
S_{BH}=4\pi\sqrt{q_{1}q_{2}}\sqrt{cd} \label{entropy}
\end{equation}
The $q_{i}$ are the charges of the system, and we would like to relate them
to the winding number and the Kaluza-Klein momentum of the system. In order to do so, we
may use Gauss's Law.
\begin{equation}
e_{1}=4\frac{mR}{16 \pi \alpha'}\frac{a}{b}\frac{16 \pi G_{10}T^{2}}{S 2 \pi R V_{5}} \hspace{5.mm}
e_{2}=\frac{n}{4 \pi R}\frac{a}{b}\frac{16 \pi G_{10}}{2\pi R V_{5} ST^2}
\end{equation}
To get the same normalization as in \cite{Sen:2005iz}, we set $\alpha'=16$,
$\frac{1}{G_{4}}=\frac{2\pi R V_{5}}{G_{10}}=\frac{1}{2}$. and
the radius of the $S^{1}$ to be $R=\sqrt{\alpha'}=4$. One then gets
\begin{equation}
q_{1}=\frac{1}{4}m \hspace{10.mm} q_{2}=\frac{1}{4}n
\end{equation}
Finally, the entropy can be expressed in terms of the winding and
the Kaluza-Klein momentum. It can then be concluded that in the
supergravity approximation, the entropy is zero. This clearly
differs from the result obtained from microscopic counting, which is
known to be
\begin{equation}
S_{BH}=2\pi\sqrt{2nm}
\label{microentropy}
\end{equation}
However, it is a well-known conjecture that one can still get a
non-vanishing macroscopic value for the entropy, if one considers
higher derivative corrections to the supergravity action
\cite{LopesCardoso:1998wt,LopesCardoso:1999xn,LopesCardoso:1999ur,Dabholkar:2004dq, Castro:2007sd, Castro:2007hc}.
The horizon area which was initially estimated to be zero, gets
``stretched'' by quantum/stringy effects. The formalism presented
above, can be used to compute the corrections to the entropy.
%%%%%%%%%%%%%%%%%%%%%%%%%%%%%%%%%%%%%%%%%%%%%%%%%%%%%%%%%%%%%%%%%%%%%%%%%%%%%%%%%%%%%%%%%%%%%%%%%%%%%%%%%%%%%%
\section{String and Brane Probes and the No-Force Conditions}
%%%%%%%%%%%%%%%%%%%%%%%%%%%%%%%%%%%%%%%%%%%%%%%%%%%%%%%%%%%%%%%%%%%%%%%%%%%%%%%%%%%%%%%%%%%%%%%%%%%%%%%%%%%%%%
\subsection{String Probes in AdS Background}
%%%%%%%%%%%%%%%%%%%%%%%%%%%%%%%%%%%%%%%%%%%%%%%%%%%%%%%%%%%%%%%%%%%%%%%%%%%%%%%%%%%%%%%%%%%%%%%%%%%%%%%%%%%%%%
The supergravity background produced by a fundamental string with
winding $w$ and momentum $p$ along a compact circle corresponds to a
two-charge black hole. Using Sen's formalism we can solve for the
values of the moduli in the near horizon limit of these black holes,
while considering the effect of higher order corrections to the
supergravity action. These solutions do not necessarily preserve
supersymmetry and we need additional checks to identify which ones
do. One possibility would be to put in a fundamental string probe in
these curved backgrounds \cite{McGreevy:2000cw} and then look for
constraints on the background ensuring the vanishing of the
force. The resulting condition might then be used as a test for
supersymmetry. It is important to note that the curvature of the
geometry is of the order of $1/\alpha'$ so $r_{AdS}/\alpha'$ is not a good
expansion parameter.
The probe solutions we are looking for have vanishing second-order
derivatives. Therefore the Nambu-Goto/DBI action corresponds to a
re-sumed series in $\alpha'$. On the other hand, we are neglecting
the backreaction of the probe strings. In the large $p,w$ limit, the
effects of back-reaction can be expanded in powers of
$1/pw$ \cite{Castro:2007sd,Castro:2007hc}. For a probe fundamental
string, the backreaction is expected to appear at the next order in
$1/pw$. As a result, the constraints that are found using these
fundamental string probes constitute a reasonable test for identifying supersymmetric
configurations. 

There is also an additional set of fundamental strings winding the
sphere in the near-horizon $AdS_{2} \times S^{2}$ geometry, which
are referred to as dual probes. These are analogous to the dual giants
considered in \cite{Grisaru:2000zn,Hashimoto:2000zp}. We will show
that these probes also yield solutions satisfying BPS-like
conditions.
%%%%%%%%%%%%%%%%%%%%%%%%%%%%%%%%%%%%%%%%%%%%%%%%%%%%%%%%%%%%%%%%%%%%%%%%%%%%%%%%%%%%%%%%%%%
\subsubsection{Probes in Poincar\'{e} Coordinates}\label{f1probes}
%%%%%%%%%%%%%%%%%%%%%%%%%%%%%%%%%%%%%%%%%%%%%%%%%%%%%%%%%%%%%%%%%%%%%%%%%%%%%%%%%%%%%%%%%%%
In Poincar\'{e} coordinates, the metric lifted to 10 dimensions is given by
\begin{equation}\label{poincare}
ds^2 = a\left(-r^2dt^2 + \frac{dr^2}{r^2}\right)+ (crdt + dy)^2 +
bd\Omega_2^2 + \sum_{i=1}^{5} dx_i^2 ,
\end{equation}
where $c$ is proportional to the momentum along the $y$-direction and
one has
\begin{equation}
B_{ty} = dr
\label{bfield}
\end{equation}
The Lagrangian of a fundamental string winding along $y$ such that
$t = \tau, y = \sigma$ is
\begin{equation}
S = \int d^{2}\sigma \left[\sqrt{-(\dot{X^2}X'^2 -
  (\dot{X}\cdot X')^2)} + B_{ty}\right]
\end{equation}
Some useful expressions are
\begin{eqnarray}\label{exp1}
\dot{X^2} &=& (c^2-a)r^2 + a\frac{\dot{r^2}}{r^2} +
b{\dot{\Omega}_{2}}^2 + \dot{x_i}^2 \nonumber \\
X'^2      &=& 1 + a\frac{r'^2}{r^2} + b\Omega_2'^2 + x_i'^2\nonumber \\
\dot{X}\cdot X'&=& a\frac{\dot{r}r'}{r^2} + cr + b
\dot{\Omega}_{2}\cdot \Omega_2' + \dot{x_i}\cdot x_i'
\end{eqnarray}
The momenta are given by
\begin{equation}
-P_a = g_{ab}\frac{X'^2 \dot{x}^b - (\dot{X}\cdot X')x^{'b}}{\sqrt{-(\dot{X^2}X'^2 -
(\dot{X}\cdot X')^2)}}
\end{equation}
For notational simplicity, we write $\sqrt{-(\dot{X^2}X'^2
-(\dot{X}\cdot X')^2)} = \sqrt{- \det G}$. We also use capital $X$
for all ten dimensional coordinates and little $x$ for the
coordinates transverse to the string world-sheet, namely $x \in
\{r,\theta, \phi, x_i\}, i = 1,2,..,5$. Writing
\begin{equation}
P_\theta \theta' + P_r r' + P_\phi \phi' +
P_{x_i}x_i' = P_x x',
\end{equation}
and using
\begin{equation}\label{id1}
\sqrt{- \det{G}} (P_\theta \theta' + P_r r' + P_\phi \phi' +
P_{x_i}x_i') - crX'^2= -(\dot{X}\cdot X') .
\end{equation}
one can show that
\begin{equation}
(-\det{G}) (P_aP_b g^{ab} + X'^aX'^b g_{ab} + (P_x x')^2) = ar^2 (X'^2)^2.
\end{equation}
which then gives a Hamiltonian
\begin{equation}
H = \sqrt{a}r\sqrt{P_a P_b g^{ab} + X'^aX'^b g_{ab} + (P_x x')^2} + cr P_x X' + B_{ty}
\end{equation}
It is observed that the terms inside the square of the Hamiltonian
can be re-written as
\begin{equation}
P_aP_b g^{ab} + X'^a X'^b g_{ab} + (P_x X')^2 = (1 \pm P_x\cdot x')^2 + (P \mp
x')_a(P \mp x')_b g^{ab},
\end{equation}
bearing in mind that $X'^2 = y'^2 + x'^2 = 1 + x'^2$. A BPS-like
solution would then be given by
\begin{equation}
P_a = x'_a.
\end{equation}
The equations of motion obtained from the Hamiltonian, assuming that
all momenta are time independent, are given by
\begin{equation}
\partial_\sigma (\sqrt{a}r+cr) P_a = \frac{\partial H}{\partial x^a}
\end{equation}
The left hand side is zero for all $r$ if $\sqrt{a}+c=0$.
To make the right hand side zero, $\theta = \pi/2$ from the $\theta$
equation. In the $r$ equation, the r.h.s. is given by
\begin{equation}
\sqrt{a}(1+P_x x') + cP_x x' + d + \frac{r^2P_r^2}{\sqrt{a}\sqrt{P_aP_b g^{ab} + X'^aX'^b
    g_{ab} + (P_x X')^2}}
\end{equation}
which vanishes for  $P_r = r'=\sqrt{a}+d=0$ from the $r$ equation.
There is then no further constraint on $P_\phi$ or $P_i$, and
immediately one sees that $|c|=|d|$. Summarizing the result, one finds
that the no-force condition for this configuration is given by
\begin{equation}
a=c^2 \qquad c=d
\label{nofcond}
\end{equation}
Substituting the solution back into the Hamiltonian, one can check that it is identically zero.
%%%%%%%%%%%%%%%%%%%%%%%%%%%%%%%%%%%%%%%%%%%%%%%%%%%%%%%%%%%%%%%%%%%%%%%%%%%%%%%%%%%%%%%%%%%%%%%%%%%%%%%%%%
\subsubsection{Probes in Global Coordinates}
%%%%%%%%%%%%%%%%%%%%%%%%%%%%%%%%%%%%%%%%%%%%%%%%%%%%%%%%%%%%%%%%%%%%%%%%%%%%%%%%%%%%%%%%%%%%%%%%%%%%%%%%%%
In global coordinates, the metric lifted to 10 dimensions is given by
\begin{equation}
ds^2 = -(1+ar^2)dt^2 + \frac{dr^2}{(1+ ar^2)}+ (crdt + dy)^2 +
bd\Omega_2^2 + \sum_{i=1}^{6} dx_i^2
\end{equation}
The metric is altered so the equations obtained in the
previous section change accordingly. Equations (\ref{exp1}) become
\begin{eqnarray}
\dot{X^2} &=& -1 + (c^2-a)r^2  + \frac{\dot{r^2}}{1+ar^2} +
b{\dot{\Omega}_{2}}^2 + \dot{x_i}^2 \nonumber \\
X'^2      &=& 1 + \frac{r'^2}{1+ar^2} + b\Omega_2'^2 + x_i'^2\nonumber \\
\dot{X}\cdot X'&=& \frac{\dot{r}r'}{1+ar^2} + cr + b\dot{\Omega}_{2}\cdot \Omega_2' + \dot{x_i}\cdot x_i'
\end{eqnarray}
The form of the identity (\ref{id1}) remains the same. The
Hamiltonian becomes now
\begin{equation}
H_g = \sqrt{1+ ar^2}\sqrt{P^2 + X'^2 + (P_x.x')^2} + crP_x x' + dr.
\end{equation}
and the equations of motion are
\begin{equation}
\partial_\sigma [(\sqrt{1+ar^2}+cr)P_x] = \frac{\partial
  H_g}{\partial x}.
\end{equation}
The $\theta$ equation again requires $\theta= \pi/2$. The r.h.s of the
$r$ equation becomes
\begin{eqnarray}
&&\frac{ar}{\sqrt{1+ar^2}}(1+P_x\cdot x') + cP_x x' + d
\nonumber \\
&&+\frac{\sqrt{1+ar^2} P_r^2 ar}{\sqrt{1+ ar^2}\sqrt{P^2 + X'^2 +
    (P_x\cdot x')^2}}
- \frac{r'^2ar^2}{2(1+ar^2)^{\frac{3}{2}}\sqrt{P^2 + X'^2 + (P_x\cdot x')^2)}}.
\end{eqnarray}
A solution would then be given by
\begin{eqnarray}
r &=&0   \nonumber\\
d+cP_x\cdot x' &=& 0 \nonumber \\
\theta &=& \frac{\pi}{2}
\end{eqnarray}
Note here that the probe is stuck at the center of the AdS space.
The same property is found in other probe solutions in similar
settings \cite{Sinha:2006sh,Sinha:2007ni,Mandal:2007ug}. The energy
of these solutions is
\begin{equation}
H_g = 1 + P_x \cdot x' = 1 -\frac{d}{c}.
\end{equation}
which vanishes identically for $c=d$.
%%%%%%%%%%%%%%%%%%%%%%%%%%%%%%%%%%%%%%%%%%%%%%%%%%%%%%%%%%%%%%%%%%%%%%%%%%%%%%%%%%%%%%%%%%%%%%%%%%%%%%%%%%%%%%%
\subsection{Dual Probes}
%%%%%%%%%%%%%%%%%%%%%%%%%%%%%%%%%%%%%%%%%%%%%%%%%%%%%%%%%%%%%%%%%%%%%%%%%%%%%%%%%%%%%%%%%%%%%%%%%%%%%%%%%%%%%%5
The derivation of the classical solutions for dual probes can be obtained
in exactly the same manner. Notice that in all the solutions that we found,
$P_\theta$ has to be zero.
%%%%%%%%%%%%%%%%%%%%%%%%%%%%%%%%%%%%%%%%%%%%%%%%%%%%%%%%%%%%%%%%%%%%%%%%%%%%%%%%%%%%%%%%%%%%%%%%%%%%%%%%%%%%%%%
\subsubsection{Dual Probes in Poincar\'{e} Coordinates}
%%%%%%%%%%%%%%%%%%%%%%%%%%%%%%%%%%%%%%%%%%%%%%%%%%%%%%%%%%%%%%%%%%%%%%%%%%%%%%%%%%%%%%%%%%%%%%%%%%%%%%%%%%%%%
The metric is just as given in (\ref{poincare}). Wrapping the dual
probe about $\phi$ and following the same procedures as in the
previous section, the Hamiltonian is now given by
\begin{equation}
H_{dualp} = \sqrt{a}r[\sqrt{(\sqrt{b}\sin \theta \pm
    \frac{P\cdot x'}{\sqrt{b}\sin\theta})^2 + (P \mp x')^2}] + crP_y +
    dr y',
\end{equation}
where now
\begin{equation}
P\cdot x' = P_rr' + P_yy' + P_\theta \theta' + P_{x_i}x_i'.
\end{equation}
A class of solutions is again given by
\begin{equation}
 P=x' .
\end{equation}
The $\theta$ equations of motion require that $\theta=\pi/2$.
The $r$ equation is
\begin{equation}
\dot{P_r} + P_r' = \sqrt{a}(\sqrt{b}\sin\theta +
\frac{P\cdot x'}{\sqrt{b}\sin\theta}) + cP_y +
\frac{\sqrt{a}r^2P_r^2}{(\sqrt{b}\sin\theta +
  \frac{P\cdot x'}{\sqrt{b}\sin\theta})} + dy'.
\end{equation}
For $\sin\theta = 1$, the r.h.s. of the above equation simplifies to
\begin{equation}
\sqrt{ab} + dy' + cP_y + P\cdot x'\sqrt{\frac{a}{b}} = 0.
\end{equation}
The no-force condition when $x'=\dot{x}=0$ is again
\begin{equation}
\sqrt{b(a-c^2)r}=0,
\end{equation}
which gives $a=c^2$ at general $r$, as we had before for
the probes.
%%%%%%%%%%%%%%%%%%%%%%%%%%%%%%%%%%%%%%%%%%%%%%%%%%%%%%%%%%%%%%%%%%%%%%%%%%%%%%%%%%%%%%%%%%%%%%%%%%%%%%%%%%%%%%%%%%%%
\subsubsection{Dual Probes in Global Coordinates}
%%%%%%%%%%%%%%%%%%%%%%%%%%%%%%%%%%%%%%%%%%%%%%%%%%%%%%%%%%%%%%%%%%%%%%%%%%%%%%%%%%%%%%%%%%%%%%%%%%%%%%%%%%%%%%%%%%%%
Again wrapping the string about $\phi$, the Hamiltonian is found to
be
\begin{equation}
H_{dualg} = \sqrt{1+ar^2}\left[\sqrt{\left(\sqrt{b}\sin \theta \pm
    \frac{P\cdot x'}{\sqrt{b}\sin\theta}\right)^2 + (P \mp x')^2}\right]
    + crP_y + dry'
\end{equation}
The $\theta$ equation requires $\theta = \pi/2$ as before. The $r$
equation gives
\begin{eqnarray}
\partial_\sigma (\sqrt{1+ar^2}P_r) &=&
\left[\frac{ar}{\sqrt{1+ar^2}}\left(\sqrt{b}\sin \theta +
    \frac{P\cdot x'}{\sqrt{b}\sin\theta}\right)+
cP_y +dy'\right]
\nonumber
\\
&+& \frac{1}{\sqrt{b}\sin\theta +
  \frac{P\cdot x'}{\sqrt{b}sin\theta}}\left[P_r^2ar -
\frac{r'^2ar}{(1+ar^2)^2}\right].
\end{eqnarray}
To make the r.h.s. zero, one demands
\begin{eqnarray}
r &=& 0 \nonumber \\
dy'+cP_y &=& 0
\end{eqnarray}
This gives $c=d$. We see that also the dual probes are stuck in the
center of AdS.
%%%%%%%%%%%%%%%%%%%%%%%%%%%%%%%%%%%%%%%%%%%%%%%%%%%%%%%%%%%%%%%%%%%%%%%%%%%%%%%%%%%%%%%%%%%%%%%%%%%%%%%%%%%%%%%%%%%%%%%
\subsection{D2 Probes in Type IIA Fundamental String Background}
%%%%%%%%%%%%%%%%%%%%%%%%%%%%%%%%%%%%%%%%%%%%%%%%%%%%%%%%%%%%%%%%%%%%%%%%%%%%%%%%%%%%%%%%%%%%%%%%%%%%%%%%%%%%%%%%%%%%%%%
In order to account for the entropy from the degeneracy of probe
brane solutions, the probes are expected to carry the same quantum
number as the background. However, the solutions we have found so
far imply relations between the momenta and the winding. As a
result, these solutions actually carry less independent charges than
the background solution and it is not expected to produce the
correct entropy upon quantization.  One could use the same
techniques as in \cite{Sinha:2007ni} to account for the entropy, and
for this the correct set of probes is required. This shall be
considered at length in the next section.

If winding fundamental strings are not carrying the correct quantum
numbers of a probe, the logical step would be to find the next
simplest candidate. This leads us to consider D2-brane probes in
type IIA theory.
%%%%%%%%%%%%%%%%%%%%%%%%%%%%%%%%%%%%%%%%%%%%%%%%%%%%%%%%%%%%%%%%%%%%%%%%%%%%%%%%%%%%%%%%%%%%%%%%%%%%%%%%%%%%%%%%%%%%%%%%%
\subsubsection{Poincar\'{e} Coordinates}
%%%%%%%%%%%%%%%%%%%%%%%%%%%%%%%%%%%%%%%%%%%%%%%%%%%%%%%%%%%%%%%%%%%%%%%%%%%%%%%%%%%%%%%%%%%%%%%%%%%%%%%%%%%%%%%%%%%%%
Now consider a D2-brane wrapping the $S^{2}$ in the same curved
background as the one given in (\ref{poincare}) in section
(\ref{f1probes}). The Lagrangian is then
\begin{equation}
L = \sqrt{\det{G+B+F}},
\end{equation}
where $G+B$ is the pull-back of the spacetime metric and NS 2-form and $F$ is the world-volume electromagnetic field.
Expanding the determinant, one can write it as
\begin{equation}
\dot{X}^2(\partial_{\theta}X)^2(\partial_{\phi}X)^2
- (\partial_{\theta}(\dot{X}\cdot \partial_{\phi}X)-\dot{X}\cdot \partial_{\phi}X))^2
+ (F\dot{X} + K_{[\phi}X_{\theta]})^2,
\end{equation}
where the squares imply implicit contraction with the metric and
\begin{eqnarray}
K_i &=& F_{ti} + B_{ty}\partial{i}y = E_i + dr\partial_{i}y \nonumber \\
F &=& F_{\theta\phi}
\end{eqnarray}
Now suppose
\begin{equation}
\dot{r}=\partial_{i}r = 0
\end{equation}
and that all second derivatives of all the fields vanish.
The $y$-momentum is given by
\begin{eqnarray}
L P_{y} &=& (b + (\partial_\theta y)^2)(b\sin^2{\theta} +
(\partial_\phi y)^2)(cr+ \dot{y})
\nonumber \\
&-& b((cr+\dot{y})(\partial_\phi y)^2 + \sin^2{\theta}(cr +
\dot{y})(\partial_\theta y)^2)
+ crF^2 + F(F\dot{y}+ K_{[\phi} y_{\theta]})
\end{eqnarray}
All terms involving a total derivative by $\tau$ or
$\phi$ are zero automatically. Since $\theta$, as a world-volume
coordinate appears explicitly in the action, we must ensure that all
terms differentiated by $\theta$ vanish.
Therefore we need only to focus our attention on setting certain
terms to zero in the equations of motion.
These terms are given by
\begin{equation}
\frac{\partial L}{\partial F}=0 \qquad
\frac{\partial L}{\partial r}=0 \qquad
\frac{\partial L}{\partial (\partial_{\theta}y)}=0.
\end{equation}
Suppose that we let $\partial_{\theta}y = \partial_{\phi}y = 0$. We can simplify the equations to
\begin{eqnarray}
\label{1}b^2\sin^2{\theta}[-ar + c(cr+\dot{y})] -2(a-c^2)rF^2 + 2cF^2\dot{y}^2 &=&0 \\
\label{2}crFE_\phi + F\dot{y}E_\phi + drE_\theta b\sin^2{\theta} &=& 0 \\
\label{3}(-a+c^2)r^2F + crF\dot{y} + crF\dot{y} + F\dot{y}^2 &=& 0
\end{eqnarray}
In eq. (\ref{1}) we need to set the piece proportional to $\sin{\theta}$ and the rest to zero independently.
Eqs. (\ref{1}) and (\ref{3}) give then $a=0$, so there is no sensible solution.
However, to obtain non-zero $P_y$ and winding (i.e. non-zero $ \partial_{i}y$),
we could equally have set $\dot{y} = 0$, but excite non-trivial electromagnetic waves.
Also, since $\theta$ is not a periodic coordinate, we can let $\partial_{\theta}y = 0$.
The equations are then reduced to
\begin{eqnarray}
\label{*1a}b(c^2-a)r &=& 0 \\
\label{*1b}r(\partial_\phi y)^2(c^2-a)
- bc^2r(\partial_\phi y)^2 + db\partial_\phi y + rF^2(c^2-a) - \label{*2a}E_\theta \partial_\phi y &=& 0\\
\label{*2b}E_\theta dr &=& 0 \\
\label{*3}K_{\phi}(crF-E_{\theta}\partial_\phi y) + drE_\theta (\partial_\phi y)^2 -dFcr^2\partial_\phi y &=& 0 \\
\label{*6}r\{(c^2-a)F - E_{\theta}\partial_\phi y\} &=& 0.
\end{eqnarray}
We certainly do not wish $r=0$ since the momentum would be identically zero. Hence,
any non-trivial solution exists only if
\begin{equation}
a = c^2.
\end{equation}
Similarly we have $E_\theta = 0.$ Then (\ref{*1a}),(\ref{*2a}) and (\ref{*3}) are satisfied.
From (\ref{*1b}) we have
\begin{equation}
\partial_\phi y [-\partial_\phi ybc^2r + db(E_\phi + dr\partial_\phi y)] = 0.
\end{equation}
Also from (\ref{*2b}) we have
\begin{equation}
crFE_\phi  = 0.
\end{equation}
Again we do not want either $r$ or $F$ to vanish for non-trivial momentum. Therefore $E_\phi = 0.$
and one gets
\begin{equation}
b(\partial_\phi y)^2[-c^2+d^2] = 0.
\end{equation}
and we have to conclude that non-trivial winding is only possible if
\begin{equation}
d^2=c^2 .
\end{equation}
Putting the solution into the Lagrangian we are left with
$\sqrt{c^2-d^2} = 0$. This suggests that the Hamiltonian would be
proportional to the momenta and thus satisfies a BPS-like condition.
%%%%%%%%%%%%%%%%%%%%%%%%%%%%%%%%%%%%%%%%%%%%%%%%%%%%%%%%%%%%%%%%%%%%%%%%%%%%%%%%%%%%%%%%%%%%%%%%%%%%%%%%%%%%%%%%%%%%%%
\subsection{Comments on Probes}
%%%%%%%%%%%%%%%%%%%%%%%%%%%%%%%%%%%%%%%%%%%%%%%%%%%%%%%%%%%%%%%%%%%%%%%%%%%%%%%%%%%%%%%%%%%%%%%%%%%%%%%%%%%%%%%%%%%%%%
The resulting no-force conditions are supposed to hold if the
background solution is to represent a supersymmetric configuration.
An interesting observation is that the background found in
\cite{Sahoo:2006pm} for the heterotic dyonic black hole also
satisfies the condition coming from the string probe analysis, which
should apply to that case as well. One should also point
out that the no-force condition will remain unchanged when
considering backgrounds of the form $AdS_{2}\times S^1\times S^{n}$, with
$n\ge 2$.

For the D2 probe, given that its mass scales as $1/g_s$ and the
attractor values of $g_s$ in the near horizon limit are proportional to
$1/\sqrt{pw}$, one could argue that they are heavy enough to produce significant
backreaction on the background. The main purpose of the exercise is
to obtain BPS conditions constraining the background, and 
one could entertain the possibility that the backreaction does not alter 
the no-force conditions we found. In fact, this is analogous to finding no-force conditions
between parallel branes, where is well known that the backreaction does not
alter the final result.  Since the constraints obtained from
the D2-brane probes are identical to those coming from fundamental strings probes,
namely $a=c^2$ and $c=d$, we argue that this is further evidence of the condition for supersymmetry,
and that backreaction from the probes will not alter the result. 
Incidentally, the first condition is required if the $AdS_2$ space were
embedded in $AdS_3$ \cite{Strominger:1998yg, Dabholkar:2006tb} \footnote{We shall review the
details in appendix A.}. If the $AdS_2$ geometry is indeed embedded in $AdS_3$,
the bulk geometry would possess the conformal symmetry of a
two-dimensional CFT, hence giving a holographic understanding of the fundamental string
worldsheet CFT. This idea has been recently explored in the literature and various WZW models 
representing the bulk geometry are have been proposed in \cite{Giveon:2006pr,Dabholkar:2007gp,Lapan:2007jx}.
In this light, we are lead to conclude that our probe solutions are probably seeing the correct physics. 
Furthermore, these solutions carry independent winding and momentum charges, and
should, upon quantization, be able to account for the black hole entropy.
%%%%%%%%%%%%%%%%%%%%%%%%%%%%%%%%%%%%%%%%%%%%%%%%%%%%%%%%%%%%%%%%%%%%%%%%%%%%%%%%%%%%%%%%%%%%%%%%%%%%%%%%%%%%%%
\section{No-Force Conditions and the WZW Action}
%%%%%%%%%%%%%%%%%%%%%%%%%%%%%%%%%%%%%%%%%%%%%%%%%%%%%%%%%%%%%%%%%%%%%%%%%%%%%%%%%%%%%%%%%%%%%%%%%%%%%%%%%%%%%%
In the previous section,  the no-force constraints we obtained
indicate that the background geometry is 
in fact an $AdS_{2}$ embedded in $AdS_{3}$. It is therefore natural to
consider the fundamental string as a 
hologram, in which the bulk
theory can be constructed as an exact CFT. This idea has been pursued
in \cite{Giveon:2006pr,Dabholkar:2007gp}. The bulk string theory is
generally formulated  as a supersymmetric $SL(2,\mathcal{R})$ 
WZW model, whose level is determined from matching the central charge
of the bulk $SL(2,\mathcal{R})$ current algebra with that of the
Virasoro algebra in the boundary CFT by virtue of the AdS/CFT
correspondence. Therefore, the type II and heterotic WZW bulk worldsheet theories 
are distinguished by their number of worldsheet supersymmetries and the central extension of the current
algebra.  The bulk worldsheet quantum effects, which can be expressed as an
expansion in $\alpha'$, together with the effects of the
worldsheet fermions, can be taken into account completely such that an
effective action can be written down \cite{Tseytlin:1993my}. The aggregate
effect is a shift in the level of the $SL(2,\mathcal{R})$ algebra.
It is then possible to consider the bosonic sector of the effective worldsheet action 
for the type II string, and determine the values of the radii correct, which should
be exact to all orders in $\alpha'$\footnote{The radii of
the $AdS$ geometries for the models under consideration is of the
order of the string scale. It is not entirely clear whether the
geometrical interpretation at such tiny length scales makes sense but this
situation is not unexpected for small black holes where \emph{a priori} we
knew that the radius of curvature is of the order of the string scale.}.

To begin with, consider the bosonic sector of a level $k$ WZW
action\footnote{Here we consider the effective action, where world-sheet quantum effects and
fermions have been accounted for by the value of the level $k$.},
which is given by 
\begin{equation}
S=\frac{k}{4\pi}\int_{\mathcal{M}} d^{2}z \mathrm{Tr}( \partial g^{-1}
\bar{\partial} g) 
+ \frac{k}{12\pi}\int_{\mathcal{N}} \mathrm{Tr}(\omega^{3})
\label{wzwaction}
\end{equation}
where $\omega=g^{-1}dg$ is the Maurer-Cartan form. As mentioned before,
we will consider the case $k=2$, in which case the bulk string theory
has the correct central 
charge. Here $g \in SL(2,R)$ and can be parametrized by
\begin{equation}
g=\left(\begin{array}{cc}
z-\frac{\gamma^{+}\gamma^{-}}{z} & \frac{\gamma^{-}}{z}    \\
-\frac{\gamma^{+}}{z}  & \frac{1}{z}\\
\end{array} \right)
\end{equation}
where $z$, $\gamma^{\pm}$ are Poincair\'{e} coordinates on $AdS_{3}$, whose metric
is given by
\begin{equation}
ds^{2}=\frac{r^{2}_{AdS_{3}}}{z^{2}}(d\gamma^{+} d\gamma^{-} +dz^{2})
\end{equation}
The action can be rewritten in terms of the Polyakov action, with
the $AdS_{3}$ metric and a Neveu-Schwarz two-form field strength
given by
\begin{equation}
H=\frac{q}{r^{3}}\eta
\label{NSWZW}
\end{equation}
and $\eta$ is the volume form associated to the $AdS_{3}$ metric. The Polyakov action reads
\begin{equation}
S=\frac{1}{2\pi\alpha'}\int d^{2}x G_{\mu\nu}\partial X^{\mu}\bar{\partial} X^{\nu} + B_{\mu\nu}\partial X^{\mu}\bar{\partial} X^{\nu}
\end{equation}
which can be rewritten in the form of (\ref{wzwaction}) as
\begin{equation}
S=\frac{1}{2\pi\alpha'} \left\{ \left(\frac{r_{AdS_{3}}^{2}}{2}\right) \int d^{2}z \mathrm{Tr}( \partial g^{-1} \bar{\partial} g)
+ \frac{q}{12} \int \mathrm{Tr}(\omega^{3}) \right\}
\end{equation}
From this expression we find that
\begin{equation}
r^{2}_{AdS_{3}}=\alpha' k \qquad \qquad  q=2\alpha'k
\end{equation}
It is well-known that $AdS_{3}$  can be reduced to $AdS_{2}\times
S^{1}$ by using an appropiate choice 
of coordinate transformations \cite{Strominger:1998yg}. The metric for
$AdS_{2}\times S^{1}$ in Poincair\'{e} coordinates is 
given by
\begin{equation}
ds^2 = a\left(-r^2dt^2 + \frac{dr^2}{r^2}\right)+ (crdt + dy)^2
\label{ads2s1metric}
\end{equation}
where
\begin{equation}
a=c^{2}=\frac{1}{4}r_{AdS_{3}}^{2}
\label{firgreat}
\end{equation}
In this set of coordinates, the field strength becomes
\begin{equation}
H_{rty}=d=\frac{q}{r_{AdS_{3}}^{3}}a
\end{equation}
Then one immediately sees that
\begin{equation}
d=\sqrt{a}
\label{secgreat}
\end{equation}
Equations (\ref{firgreat}) and (\ref{secgreat}) are the same as the
no-force constraints found in section 4. The key point is that
these conditions directly result from the symmetry of the WZW model
and are independent of $\alpha'$ corrections, and the probes in
earlier sections constrain the background to have precisely this
symmetry using only no-force arguments.  

Setting $\alpha'=16$ in our convention, and considering that for  
type II strings, $k=2$ \cite{Giveon:2006pr,Dabholkar:2007gp}, one gets  
\begin{equation}
a=8 \qquad \qquad c=d=2\sqrt{2}
\end{equation}
which when substituted into (\ref{entropy}), reproduces the result in
(\ref{microentropy}). This result reinforces the fact that Sen's
entropy formalism, which corresponds to the Wald's formula, is consistent
with the AdS/CFT construction.
%%%%%%%%%%%%%%%%%%%%%%%%%%%%%%%%%%%%%%%%%%%%%%%%%%%%%%%%%%%%%%%%%%%%%%%%%%%%%%%%%%%%%%%%%%%%%%%%%%%%%%%%%%%%%%%%%%%%%%%
\section{Higher Derivative Corrections}
%%%%%%%%%%%%%%%%%%%%%%%%%%%%%%%%%%%%%%%%%%%%%%%%%%%%%%%%%%%%%%%%%%%%%%%%%%%%%%%%%%%%%%%%%%%%%%%%%%%%%%%%%%%%%%%%%%%%%%%
In the previous section we determined the geometry by considering a
general WZW model proposed in  \cite{Giveon:2006pr,Dabholkar:2007gp}
. However, it is still of interest to explore the effects of higher
derivative corrections to the entropy coming from leading $\alpha'$
corrections to the IIB supergravity action and investigate the role
played by the no-force constraints and the field redefinition
parameters. The correction terms in the action can be determined
from string scattering amplitudes
\cite{Gross:1986iv,Grisaru:1986vi}. Coefficients for terms
proportional to the Ricci tensor remain undetermined, since the
scattering amplitudes are evaluated on-shell (to lowest order in
$\alpha'$), and so they vanish. Actions with different linear
combinations of these terms are related by field redefinitions
\cite{Tseytlin:1986zz}, since these do not alter the scattering
amplitudes. However, while remaining ambiguous, they would
contribute to the entropy via Sen's entropy function
\cite{Wald:1993nt} as discussed earlier. These ambiguities are
expected to be canceled out once all corrections from all orders in
$\alpha'$ are taken into account. Therefore,  it is interesting to
see how such ambiguities enter the determination of the entropy at
finite order, and if one could find a special field redefinition
that gives the full entropy at this order without receiving further
corrections. It would also be gratifying to see that higher order
corrections at some finite order in $\alpha'$ do give rise to a
stretched horizon in general, and to see whether any of the
solutions stretching the horizon are supersymmetric at that order.

As a first step we will consider the effect of adding general
$\mathcal{R}^4$ terms to the solutions of the attractor equations (\ref{attractor}).
We refer the reader to appendix B for further details.
We start by writing down all possible Lorentz scalars that can be
constructed from contraction of four Riemann tensors. For $d>8$, one
has 26 linearly independent Lorentz scalars \cite{Fulling:1992vm}. 
Many of these terms are set to zero when considering the on-shell
supergravity action. However, we will consider a general linear
combination of these terms, so the next-to-leading order Lagrangian
is written as
\begin{equation}
L_{off-shell}=\frac{1}{8}\zeta(3)(\alpha')^3\sum^{26}_{i=1}a_{i}\mathcal{R}^{4}_{i}
\label{corralpha}
\end{equation}
We now use this form of the Lagrangian to construct the corrections to the entropy
function in four dimensions. Naively, it would seem like an
ill-advised idea, since most of the coefficients are undetermined.
However, we will see that many of these contractions give the same
answer for symmetric spaces.

To begin with we consider the dimensionally reduced action ignoring
the Kaluza-Klein flux and evaluate the Lagrangian with the
four-dimensional near horizon metric (\ref{metric1}). We will then
consider more general near-horizon geometries, but still restricting
to corrections in the lower-dimensional gravity sector. Finally in
section 6.2, we will lift the near horizon $AdS_2 \times S^2 $
metric to ten dimensions and evaluate the Lagrangian including the
contributions from the NS flux. This is equivalent to considering
all the Kaluza-Klein and NS fluxes when dimensionally reducing the
action to four dimensions.
%%%%%%%%%%%%%%%%%%%%%%%%%%%%%%%%%%%%%%%%%%%%%%%%%%%%%%%%%%%%%%%%%%%%%%%%%%%%%%%%%%%%%%%%%%%%%%%%%%%%%%%%%%%%%%%%%%%
\subsection{$AdS_{2}\times S^{n}$ Horizon}
%%%%%%%%%%%%%%%%%%%%%%%%%%%%%%%%%%%%%%%%%%%%%%%%%%%%%%%%%%%%%%%%%%%%%%%%%%%%%%%%%%%%%%%%%%%%%%%%%%%%%%%%%%%%%%%%%%%
Consider a four dimensional black
hole with near horizon geometry $AdS_{2}\times S^{2}$
\begin{equation}
ds^{2}=a\left(-r^{2}dt^2+\frac{dr^2}{r^2}\right)+b(d\theta^2+\sin^2\theta d\phi^2)
\end{equation}
The entropy function receives higher derivative contributions of the form
\begin{equation}
f(a,b,c,d)=\frac{1}{4}a b S \left[ -\frac{1}{a}+\frac{1}{b}+\frac{c^2}{4a^2}+\frac{d^2}{4a^2}\right]
+\frac{1}{8}\zeta(3)(\alpha')^3 S \left[\sum_{i=1}^{26}a_{i}\mathcal{R}^{4}_{i}(a,b,c,d)\right]
\end{equation}
Given that one is free to redefine the fields, and given that some contractions
yield the same dependence on the radii, it is simpler to reduce the number of field
redefinition parameters to perform the analysis. This is described in more detail in the appendix.
Here it suffices to say that the corrections will only include four undefined parameters
that remain arbitrary and can be chosen in any way. 

Using then the lagrangian (\ref{corralpha}), one can obtain the 
attractor equations (\ref{attractor}). The equation $c\frac{\partial g}{\partial
  c}=d\frac{\partial g}{\partial d}$ will immediately yield
$c=d$. The other equations are more complicated, and therefore are included in the appendix.  
We can restrict our search to supersymmetric configurations by imposing
the no-force condition we derived from the probe analysis, namely,
$a=c^2$ and $c=d$, the latter of which has
been automatically implemented by the equations of motion. The system we are left with is
over-determined for given values of the parameters
and in general has \emph{no} solution, which suggests that the
no-force condition is \emph{incompatible} with the attractor equations
obtained from the truncated action excluding flux terms. This strongly
indicates that the truncation is inconsistent with supersymmetry.

One could consider more general horizon geometries. We will keep the
$AdS_{2}$ factor given that the black hole is supposed to be
extremal, and allow the dimension of the sphere $n$ to lie between 3
and 8, and proceed in the same manner as before. Again, the number
of independent scalars formed with four Riemann tensors is 26, and
once more we will ignore corrections involving the Kaluza-Klein and NS
fluxes. It is straightforward to compute the contractions, given
that we are dealing only with the gravity sector and the geometry is that 
of maximally symmetric spaces. The analysis proceeds as before, and one finds
that the resulting attractor equations are again inconsistent with the
no-force conditions. Ignoring issues of consistency for the moment, we could
try to look for special regions in the space of redefinition
parameters where the no-force conditions and the attractor equations
are actually consistent with each other. This is achieved by treating the 
field redefinition parameters as variables.
Let us first eliminate two parameters, so that we are left with an equation whose
dependence on the remaining ones drops out
\begin{eqnarray}
&&\frac{1}{a^2bn(n-1)(n-2)}\left[3a^6n^2+3a^6n^4-6a^6n^3+2a^4bc^2n^2-2a^4bc^2n+36ab^5\right.
\nonumber
\\&&\left.-18a^2b^4n^2-12b^5c^2-6a^5bn^2
+6a^5bn+18a^2b^4n\right]=0
\end{eqnarray}
Let us now impose the no-force conditions and find solutions for $a$ in terms of $b$. The only real solution
to the system is given by
\begin{equation}
a=\frac{4}{3}\frac{b}{n(n-1)}
\end{equation}
with the other roots being imaginary. The other remaining pair of equations can be used to fix $b$ in terms
of the parameters, and solutions could be found explicitly by tuning the parameters. However, we should stress
that it seems to be inconsistent to use the no-force conditions when truncating the action
to the gravity sector, so let us now consider the effects of using the no-force constraints
when taking into account the fluxes.
%%%%%%%%%%%%%%%%%%%%%%%%%%%%%%%%%%%%%%%%%%%%%%%%%%%%%%%%%%%%%%%%%%%%%%%%%%%%%%%%%%%%%%%%%%%%%%%%%%%%%%%%%%%%%%%%%%%%%%%%%5
\subsection{Adding the Fluxes}
%%%%%%%%%%%%%%%%%%%%%%%%%%%%%%%%%%%%%%%%%%%%%%%%%%%%%%%%%%%%%%%%%%%%%%%%%%%%%%%%%%%%%%%%%%%%%%%%%%%%%%%%%%%%%%%%%%%%%%%%%%%
In the previous subsection we concluded that the gravity sector alone in
general does not stretch the horizon, unless one
restricts the field redefinition parameters to a special subspace. It
is unclear whether the procedure is consistent. Given that we have
neglected the flux terms in the first place, one
should after all look for solutions in their presence. In
principle, this makes the calculation much more involved, and one could hope to find a solution only
if there were a simplification upon the use of the no-force condition.

We start by lifting the near-horizon $AdS_2 \times S^2$ metric to ten
dimensions as in (\ref{poincare}). The resulting geometry is that
of $AdS_2 \times S^2 \times S^1 \times T^{5} $. In order to take the NS
fluxes into account, we will consider the
conjecture in \cite{Gross:1986mw,Kehagias:1997cq},
which gives a prescription for the eight derivative terms in the action, involving
the NS-NS form. It should be noted that there are further corrections involving
fluxes to the tree-level NS-NS sector at the $\alpha'^3$ order \cite{Peeters:2001ub}. For instance, terms of the 
form $\mathcal{R}^{3}H^{2}$ are present, but are not reproduced when writing the NS-NS form as a torsion
\footnote{We thank M. B. Green, P. Vanhove and D. Tsimpis for their remarks on this issue.}.
Indeed, the prescription we use here is only valid for the four-point contribution \cite{Policastro:2006vt} 
to the effective action, but nevertheless, the major simplification that follows indicates 
that these terms are consistent with supersymmetry.

The way to implement this conjecture, is by
defining an effective Riemann curvature such that the flux terms are
packaged in certain combinations carrying the
symmetries of the Riemann tensor. Since the derivatives of all the
fields are assumed to vanish in the near horizon limit, the
surviving terms are given by
\begin{equation}
{\tilde{R}_{rs}}\ ^{pq}={R_{rs}}^{pq}+{\nabla_{[r}H_{s]}}^{pq}-{H_{[r}}^{u[p}{H_{s]u}}^{q]}
\end{equation}
Corrections at order $\alpha'^3$ are then built by contracting this
generalized Riemann curvature. The first term we shall consider, is
the well-known $C^4$ Weyl curvature term. From the superspace formalism of type
IIB string theory \cite{Peeters:2000qj, Green:1998by, Green:2005qr}, one can show that the combination of eight
derivative terms, that will contribute to the action at this order
is given by
\begin{equation}
\tilde{C}^{4}=-\frac{1}{4}\tilde{C}_{pqrs}{\tilde{C}_{pq}}\ ^{tu}{\tilde{C}_{rt}}\ ^{vw}\tilde{C}_{suvw}
+\tilde{C}^{pqrs}{{{\tilde{C}_{p}}\ ^{t}}_{r}}\ ^{u}{{{\tilde{C}_{t}}\ ^{v}}_{q}}\ ^{w}\tilde{C}_{uvsw},
\label{weyl4}
\end{equation}
where $\tilde{C}$ is the generalised Weyl tensor obtained from $\tilde{R}$.
The action can now be evaluated for the metric (\ref{poincare}) and
NS field (\ref{bfield}). These terms will give rise to all the
Kaluza-Klein fields that will contribute to the entropy function of
the four dimensional black hole. We will not attempt to write down
the explicit form of the entropy function or of the attractor
equations for this case, given their formidable length, but in
principle one could try to solve them numerically. Instead of
proceeding in this way, we will impose the no-force conditions found
in section 4, as mentioned above.

In the previous cases, we saw that the fourth equation in (\ref{attractor})
always implied the condition $c=d$. This is no longer true, given that
the additional terms, do not preserve the symmetry between $c$ and
$d$. However from the probe analysis, we found that for the
existence of a supersymmetric solution, the background
was required to satisfy $c=d$ in addition to $a=c^2$. Remarkably, by demanding
these conditions, one finds that the fourth attractor equation in (\ref{attractor}) is
automatically satisfied, whereas the remaining equations are
simplified tremendously.

At this point, we are left in principle with three equations and two unknowns, namely $\{b,d\}$. However it turns
out that the first two equations become degenerate, so one has two
equations and two unknowns that can be solved
numerically. These are
\begin{eqnarray}
&&54a^4b^3-27a^3b^4-40\zeta(3)ab^3 -320\zeta(3)a^3b
+240\zeta(3)a^2b^2+35\zeta(3)b^4+4480\zeta(3)a^4=0
\nonumber \\
&&27a^3b^4-640\zeta(3)a^3b-35\zeta(3)b^4+240\zeta(3)a^2b^2
+13440\zeta(3)a^4=0
\end{eqnarray}
One can check that $\{a = 1.15093, b = -23.11745\}$ is a numerical root to this system. However, this
solution has negative $b$, so the signature of the horizon is changed. It was pointed out to us that a change in
signature might represent a non-geometric background, and therefore should not be regarded as being physical
\footnote{We thank A. Castro for useful discussions regarding this issue} so we are left with the
conclusion that there is no supersymmetric solution that can be found with the $C^4$ Weyl curvature term.

We could still make the same arguments as before, and look for more generic eight derivative terms. Bearing in
mind that perturbative string calculations do not determine the coefficients of these terms uniquely, there is no
reason why they should be excluded \emph{a priori}. Therefore, we can proceed as in the previous subsections and
consider a linear combination of all 26 $\mathcal{R}^4$ scalars, evaluating this for the generalized Riemann
curvature so to include the fluxes. The equations are of course, much
more complicated, but fortunately it is still true that by
imposing the no-force conditions above, the last equation in (\ref{attractor}) is satisfied, 
and the first two equations become identical. Hence, unlike the previous case where the corrections from
the fluxes are ignored, one is \emph{always} left with two equations relating
$d$ and $b$ for any choice of field-redefinition parameters, and we see that in this case the truncation of the action is \emph{consistent} with the no-force constraints.
However, there is no guarantee that the solutions obtained from the present truncation are physical, as
we showed explicitly for the Weyl curvature correction term. The value
of the entropy will depend on the arbitrary parameters and the dependence is not expected
to drop out at any finite order in $\alpha'$. Our analysis in this section supports the special
status of the no-force condition and reinforces the need to study the system using an exact CFT.
%%%%%%%%%%%%%%%%%%%%%%%%%%%%%%%%%%%%%%%%%%%%%%%%%%%%%%%%%%%%%%%%%%%%%%%%%%%%%%%%%%%%%%%%%%%%%%%%%%%%%%%%%%%%%%5
\section{Conclusions}
%%%%%%%%%%%%%%%%%%%%%%%%%%%%%%%%%%%%%%%%%%%%%%%%%%%%%%%%%%%%%%%%%%%%%%%%%%%%%%%%%%%%%%%%%%%%%%%%%%%%%%%%%%%%%%%
In this paper we have studied the stretching of the horizon of the type II
fundamental string small black hole, and analysed
the conditions for preserving supersymmetry.
One issue we addressed was how to identify supersymmetric configurations
after taking into account higher derivative terms in the action. Given that the $\alpha'$ corrected
Killing spinor equations are yet to be determined, one needs an
alternative formalism for identifying supersymmetric configurations.
Our procedure has been to probe the background with strings and
branes, and find constraints that lead to a vanishing force.
It is well known that extremal four dimensional black holes have
$AdS_2\times S^2$ as their near horizon geometry. It has, in the past been
assumed that dimensionally reduced small black holes arising from wrapped strings also
have this geometry in the presence of $\alpha'$
corrections. Therefore, we considered solutions of probe
strings/branes moving in an $AdS_2\times S^2$ background,
and determined constraints relating the radius of the $AdS_2$ to the fluxes.
The ``no-force'' condition $a=c^2$ we found coincides with the
condition that enhances the $AdS_{2}$ symmetry to
$AdS_{3}$.

Motivated by this fact we then looked at the formulation of the bulk geometry in
terms of an $SL(2,\mathcal{R})$ WZW model which takes into account all
$\alpha'$ corrections.  We determined the values of the moduli by considering the
models constructed in \cite{Giveon:2006pr,Dabholkar:2007gp} and evaluated the entropy
using eq. (\ref{entropy}), which matches the result obtained from microscopic counting. The WZW action
also implies both no-force conditions we determined using the probes. This gives further evidence
supporting the level two $SL(2,\mathcal{R})$ WZW model as the
holographic dual to the fundamental string worldsheet. It is however surprising
that the probe analysis does not seem to receive any higher derivative corrections,
contrary to naive expectations.

We also considered
higher derivative corrections to the horizon employing the entropy
function formalism while imposing the no-force conditions on the
background.  We showed that upon inclusion of all
possible $R^{4}$ terms and ignoring corrections to the NS and
Kaluza-Klein fluxes, the equations are inconsistent with
supersymmetry.

Therefore we then studied the role played by the corrections
involving the fluxes in stretching the horizon. The fluxes are taken
into account by lifting the near horizon metric to ten dimensions
and using the prescriptions in \cite{Gross:1986mw,Kehagias:1997cq}.
We showed that the no-force conditions are automatically consistent
with the attractor equations for any field redefinition. This
resulted from the tremendous simplification obtained after imposing
them. Solutions of the attractor equations and subsequently the
entropy depend, however, on the ambiguous field redefinition
parameters. In particular, motivated by on-shell superspace
arguments, we studied corrections taking the form of the well-known
$C^4$ Weyl curvature term. We found a numerical root to the
attractor equations where the signature of the near-horizon geometry
is changed. We conclude that there are no physically acceptable
solutions for this particular choice of field redefinition
parameters, but others will in general give solutions in which the
horizon is stretched.

It is clear from our analysis that it is unlikely that higher
derivative corrections at finite order can possibly reproduce
the full microscopic entropy of the type II small black holes.
This is different from the heterotic string where the microscopic entropy can
be obtained by including only the four-derivative corrections, namely
the Gauss-Bonnet term \cite{Sen:2005kj}. It has been argued that this
is the case because, in the heterotic case, there are non-renormalisation theorems at work.
This is justified by studying the same black hole in five dimensions and assuming
that the near horizon geometry has an $AdS_3$ factor. The anomalies of the boundary CFT can
be determined uniquely from certain terms in the bulk, which in turn
fix the value of the entropy \cite{Kraus:2005vz,Kraus:2005zm}. The
same type of argument fails in type II theories given the absence of anomalies, which suggests
that one must consider all $\alpha'$ corrections to the
supergravity action. This is also consistent with the fact
that the field redefinition parameters do not drop out, at least
at this order and possibly at any finite order in $\alpha'$.  The WZW formulation of the
bulk geometry is probably the best handle in understanding these small black holes.
%%%%%%%%%%%%%%%%%%%%%%%%%%%%%%%%%%%%%%%%%%%%%%%%%%%%%%%%%%%%%%%%%%%%%%%%%%%%%%%%%%%%%%%%%%%%%%%%%%%%%%
\appendix
%%%%%%%%%%%%%%%%%%%%%%%%%%%%%%%%%%%%%%%%%%%%%%%%%%%%%%%%%%%%%%%%%%%%%%%%%%%%%%%%%%%%%%%%%%%%%%%%%%%%%%
\section{$AdS_2$ and $AdS_3$}
%%%%%%%%%%%%%%%%%%%%%%%%%%%%%%%%%%%%%%%%%%%%%%%%%%%%%%%%%%%%%%%%%%%%%%%%%%%%%%%%%%%%%%%%%%%%%%%%%%%%%%
Here we show explicitely how to reduce the $AdS_{3}$ metric to $AdS_{2}\times S^{1}$,
following \cite{Strominger:1998yg}.

Consider the $AdS_3$ metric in Poincar\'{e} coordinates.
\begin{equation}
ds_3^2 = \frac{1}{y^2}(dw^+dw^- + dy^2).
\end{equation}
Now apply the following transformations
\begin{eqnarray}
w^- &=& t^-, \nonumber \\
w^+ &=& \frac{1}{2}\exp{(2u)}, \nonumber \\
y &=& \sqrt{\frac{t^+-t^-}{T}}\exp{(u)},
\end{eqnarray}
for some real parameter $T$. The resultant metric is then
\begin{equation}
ds_3^2 = \frac{(dt^+-dt^-)^2}{4(t^+-t^-)^2}+ \frac{du(dt^++
  dt^-)}{(t^+-t^-)}+ du^2.
\end{equation}
Rewriting $2t= t^++t^-$ and $2x = t^+-t^-$, the metric becomes
\begin{equation}
ds_3^2 = (\frac{-dt^2+dx^2}{4x^2}) + (\frac{dt}{2x}+ du)^2.
\end{equation}
This metric is then identified with that of $AdS_2 \times S^1$, with $a=c^2$ using the
parametrisation given in section (\ref{f1probes}).
%%%%%%%%%%%%%%%%%%%%%%%%%%%%%%%%%%%%%%%%%%%%%%%%%%%%%%%%%%%%%%%%%%%%%%%%%%%%%%%%%%%%%%%%%%%%%%%%%%%%%%
\section{$\mathcal{R}^{4}$ Corrections}
%%%%%%%%%%%%%%%%%%%%%%%%%%%%%%%%%%%%%%%%%%%%%%%%%%%%%%%%%%%%%%%%%%%%%%%%%%%%%%%%%%%%%%%%%%%%%%%%%%%%%%
In this appendix we include some of the details regarding the computation of $\mathcal{R}^{4}$
corrections to the entropy of the type II fundamental string black hole. 

General $\mathcal{R}^{4}$ terms, for $d>8$, can be expressed in terms of a basis of 26 
independent Lorentz scalars built from contractions of four Riemann tensors. A particular 
basis for these is reproduced below \cite{Fulling:1992vm} 
\begin{center}
\begin{tabular}{ll}
$\mathcal{R}^{4}_{1}=R^{4}$&$\mathcal{R}^{4}_{14}=R^{pq}R^{rs}{{{R^{t}}_{p}}^{u}}_{r}R_{tqus}$\\
$\mathcal{R}^{4}_{2}=R^{2}R^{pq}R_{pq}$&$\mathcal{R}^{4}_{15}=RR^{pqrs}{R_{pq}}^{tu}R_{rstu}$\\
$\mathcal{R}^{4}_{3}=RR^{pq}{R_{p}}^{r}R_{qr}$&$\mathcal{R}^{4}_{16}=RR^{pqrs}{{{R_{p}}^{t}}_{r}}^{u}R_{qtsu}$\\
$\mathcal{R}^{4}_{4}=(R^{pq}R_{pq})^2$&$\mathcal{R}^{4}_{17}=R^{pq}{{{R_{p}}^{r}}_{q}}^{s}{R^{tuv}}_{r}R_{tuvs}$\\
$\mathcal{R}^{4}_{5}=R^{pq}{R_{p}}^{r}{R_{q}}^{s}R_{rs}$&$\mathcal{R}^{4}_{18}=R^{pq}R^{rstu}{{R_{rs}}^{v}}_{p}R_{tuvq}$\\
$\mathcal{R}^{4}_{6}=RR^{pq}R^{rs}R_{prqs}$&$\mathcal{R}^{4}_{19}=R^{pq}R^{rstu}{{R_{r}}^{v}}_{tp}R_{svuq}$\\
$\mathcal{R}^{4}_{7}=R^{pq}R^{rs}{R_{r}}^{t}R_{psqt}$&$\mathcal{R}^{4}_{20}=(R^{pqrs}R_{pqrs})^2$\\
$\mathcal{R}^{4}_{8}=R^2R^{pqrs}R_{pqrs}$&$\mathcal{R}^{4}_{21}=R^{pqrs}{R^{pqr}}^{t}{R^{uvw}}_{s}R_{uvwt}$\\
$\mathcal{R}^{4}_{9}=RR^{pq}{R^{rst}}_{p}R_{rstq}$&$\mathcal{R}^{4}_{22}=R^{pqrs}{R_{pq}}^{tu}{R_{tu}}^{vw}R_{rsvw}$\\
$\mathcal{R}^{4}_{10}=R^{pq}R_{pq}R^{rstu}R_{rstu}$&$\mathcal{R}^{4}_{23}=R^{pqrs}{R_{pq}}^{tu}{R_{rt}}^{vw}R_{suvw}$\\
$\mathcal{R}^{4}_{11}=R^{pq}{R_{p}}^{r}{R^{stu}}_{q}R_{stur}$&$\mathcal{R}^{4}_{24}
=R^{pqrs}{R_{pq}}^{tu}{{{R_{r}}^{v}}_{t}}^{w}R_{svuw}$\\
$\mathcal{R}^{4}_{12}=R^{pq}R^{rs}{R^{tu}}_{pr}R_{tuqs}$&$\mathcal{R}^{4}_{25}
=R^{pqrs}{{{R_{p}}^{t}}_{r}}^{u}{{{R_{t}}^{v}}_{u}}^{w}R_{qvsw}$\\
$\mathcal{R}^{4}_{13}=R^{pq}R^{rs}{{{R^{t}}_{p}}^{u}}_{q}R_{trus}$&$\mathcal{R}^{4}_{26}
=R^{pqrs}{{{R_{p}}^{t}}_{r}}^{u}{{{R_{t}}^{v}}_{q}}^{w}R_{uvsw}$
\end{tabular}
\end{center}
Corrections to the supergravity lagrangian can be computed by considering linear combinations of the above 
$\mathcal{R}^4$ terms. Let us first evaluate the corrections coming from the gravitational sector only.
This is, one ignores contributions coming from the fluxes. One can check that all the contractions 
are proportional to six different combinations
\begin{center}
\begin{tabular}{ll}
$\mathcal{R}^{4}_{1}$&$\sim \frac{(a-b)^4}{a^{4}b^{4}}$\\
$\mathcal{R}^{4}_{2},\mathcal{R}^{4}_{8}$&$\sim\frac{(a-b)^{2}(a^{2}+b^{2})}{a^{4}b^{4}}$\\
$\mathcal{R}^{4}_{3},\mathcal{R}^{4}_{6},\mathcal{R}^{4}_{9},\mathcal{R}^{4}_{15}$&$\sim\frac{(a-b)(a^3-b^3)}{a^{4}b^{4}}$\\
$\mathcal{R}^{4}_{4},\mathcal{R}^{4}_{10},\mathcal{R}^{4}_{20}$&$\sim\frac{(a^{2}+b^{2})^{2}}{a^{4}b^{4}}$\\
$\mathcal{R}^{4}_{5},\mathcal{R}^{4}_{7},\mathcal{R}^{4}_{11},\mathcal{R}^{4}_{12},\mathcal{R}^{4}_{13},\mathcal{R}^{4}_{14},
\mathcal{R}^{4}_{17},\mathcal{R}^{4}_{18},\mathcal{R}^{4}_{21},\mathcal{R}^{4}_{22},\mathcal{R}^{4}_{23},\mathcal{R}^{4}_{25},
\mathcal{R}^{4}_{26}$&$\sim \frac{(a^{4}+b^{4})}{a^{4}b^{4}}$\\
$\mathcal{R}^{4}_{16},\mathcal{R}^{4}_{19},\mathcal{R}^{4}_{24}$&$=0$
\end{tabular}
\end{center}
Given that one is free to redefine the fields, we consider a generic linear combination (eq. (\ref{corralpha}))
and group terms that are proportional to the same function of $(a,b)$, the radii of the geometry. 
For instance, $\mathcal{R}^{4}_{2}$ and $\mathcal{R}^{4}_{8}$ give the same result, up to a factor of 2, so we set 
\begin{eqnarray}
a_{2}\mathcal{R}^{4}_{2}+a_{8}\mathcal{R}^{4}_{8}&=&(a_{2}+2a_{8})\mathcal{R}_{2}^{4}\nonumber \\
                                                 &=&\tilde{a}_{2}\mathcal{R}_{2}^{4} \nonumber
\end{eqnarray}
One can proceed analogously with the remaining terms. In the end, one is left with five parameters, namely,
$\tilde{a}_{1}$, $\tilde{a}_{2}$, $\tilde{a}_{3}$, $\tilde{a}_{4}$ and $\tilde{a}_{5}$. The last parameter, 
$\tilde{a}_5$, is fixed from an on-shell string amplitude calculation \cite{Grisaru:1986vi}. 
The rest remain arbitrary. One can now obtain the attractor equations
(\ref{attractor}). The last equation immediately gives $c=d$. Substituing this in the first three equations
yields the expressions 
\begin{eqnarray}
-{a}^{3}{b}^{4}+{a}^{4}{b}^{3}+\frac{1}{2}{a}^{2}{b}^{4}{c}^{2}+\zeta(3)\left[22528 a_{3}{a}^{4} +10240
a_{2}{a}^{4} +4096a_{1}{a}^{4}-16384a_{1}{b}^{3}a \right.
\nonumber \\
-16384 a_{1}b{a}^{3}+24576 a_{1}{b}^{2}{a}^{2}+ 43008 a_{4}{b}^{2}{a}^{2}-20480a_{2}{b}^{3}a
+20480 a_{2}{b}^{2}{a}^{2} \nonumber \\
-20480 a_{2}b{a}^{3}-22528 a_{3}{b}^{3}a-22528
a_{3}b{a}^{3}+21504 a_{4}{b}^{4}+21504 a_{4}{a}^{4} \nonumber \\
\left.+68608 a_{5}{b}^{4}+68608
a_{5}{a}^{4}+22528 a_{3}{b}^{4}+10240 a_{2}{b}^{4}+4096 a_{1}{b}^{4}\right]=0 
\nonumber \\
\end{eqnarray}
\begin{eqnarray}
{a}^{4}{b}^{3}-\frac{1}{2}{a}^{2}{b}^{4}{c}^{2}+\zeta(3)\left[-24576 a_{1}{b}^{2}{a}^{2}+32768 a_{1}{b}^{3}a
-20480 a_{2}{b}^{2}{a}^{2}+40960 a_{2}{b}^{3}a\right.
\nonumber \\
-43008 a_{4}{b}^{2}{a}^{2}+45056 a_{3}{b}^{3}a-205824 a_{5}{b}^{4}+21504 a_{4}{a}^{4}
-64512 a_{4}{b}^{4}+4096 a_{1}{a}^{4}
\nonumber \\
\left.+10240 a_{2}{a}^{4}-30720 a_{2}{b}^{4}-67584 a_{3}{b}^{4}+68608 a_{5}{a}^{4}-12288 a_{1}{b}^{4}+22528 a_{3}{a}^{4}\right]=0
\nonumber \\
\end{eqnarray}
\begin{eqnarray}
a^{3}b^{4}-\frac{1}{2}a^{2}b^{4}c^{2}+\zeta(3)\left[24576 a_{1}b^{2}a^{2}-32768 a_{1}b a^{3}
+20480 a_{2}b^{2}a^{2}+43008 a_{4}b^{2}a^{2}
\right.\nonumber \\
-45056 a_{3}b a^{3}-40960 a_{2} b a^{3}-68608 a_{5}b^{4}+64512 a_{4}a^{4}-21504 a_{4}b^{4}+12288 a_{1}a^{4}
\nonumber \\
\left.+30720 a_{2}a^{4}-10240 a_{2}b^{4}-22528 a_{3}b^{4}+205824 a_{5}a^{4}-4096 a_{1}b^{4}
+67584 a_{3}a^{4}\right]=0
\nonumber \\
\end{eqnarray}
where we have dropped the tilde to avoid cluttering. 

More general geometries can also be considered. For horizon geometries
of the form $AdS_{2}\times S^{n}$, with the dimension of the sphere between 3
and 8, the $\mathcal{R}^{4}$ contractions are again easy to compute
\begin{center}
\begin{tabular}{ll}
$\mathcal{R}^{4}_{1}$&$\sim \left(\frac{n(n-1)}{b}-\frac{2}{a}\right)^{4}$\\
$\mathcal{R}^{4}_{2},\mathcal{R}^{4}_{9}$&$\sim\left(\frac{n(n-1)}{b}-\frac{2}{a}\right)^2\left(\frac{n(n-1)^2}{b^2}+\frac{2}{a^2}\right)$\\
$\mathcal{R}^{4}_{3},\mathcal{R}^{4}_{6}$&$\sim\left(\frac{n(n-1)}{b}-\frac{2}{a}\right)\left(\frac{n(n-1)^3}{b^3}-\frac{2}{a^3}\right)$ \\
$\mathcal{R}^{4}_{4}$&$\sim\left(\frac{n(n-1)^2}{b^2}+\frac{2}{a^2}\right)^2$\\
$\mathcal{R}^{4}_{5},\mathcal{R}^{4}_{7}$&$\sim \left(\frac{n(n-1)^4}{b^4}+\frac{2}{a^4}\right)$ \\
$\mathcal{R}^{4}_{8}$&$\sim \left(\frac{n(n-1)}{b}-\frac{2}{a}\right)^2\left(\frac{n(n-1)}{b^2}+\frac{2}{a^2}\right)  $ \\
$\mathcal{R}^{4}_{10}$&$\sim\left(\frac{n(n-1)}{b^2}+\frac{2}{a^2}\right)\left(\frac{n(n-1)^2}{b^2}+\frac{2}{a^2}\right)$\\
$\mathcal{R}^{4}_{11},\mathcal{R}^{4}_{12},\mathcal{R}^{4}_{17},$&$\sim \left(\frac{n(n-1)^3}{b^4}+\frac{2}{a^4}\right)$ \\
$\mathcal{R}^{4}_{14}$&$\sim  \left((n^2+n(n-2))\frac{(n-1)^2}{b^4}+\frac{4}{a^4}\right) $ \\
$\mathcal{R}^{4}_{15}$&$\sim\left(\frac{n(n-1)}{b}-\frac{2}{a}\right)\left(\frac{n(n-1)}{b^3}-\frac{2}{a^3}\right)$ \\
$\mathcal{R}^{4}_{16},\mathcal{R}^{4}_{19},\mathcal{R}^{4}_{24}$&$\sim \frac{1}{b^4}$\\
$\mathcal{R}^{4}_{18},\mathcal{R}^{4}_{21}$&$\sim \left(\frac{n(n-1)^2}{b^4}+\frac{2}{a^4}\right)$\\
$\mathcal{R}^{4}_{20}$&$\sim\left(\frac{n(n-1)}{b^2}+\frac{2}{a^2}\right)^2$ \\
\end{tabular}

\begin{tabular}{ll}
$\mathcal{R}^{4}_{22},\mathcal{R}^{4}_{23}$&$\sim\left(\frac{n(n-1)}{b^4}+\frac{2}{a^4}\right)$ \\
$\mathcal{R}^{4}_{25}$&$\sim \frac{n^2+2n(n-2)+n^2(n-2)^2}{b^4}+\frac{4}{a^4}$  \\
$\mathcal{R}^{4}_{26}$&$\sim \frac{n+2n^2(n-2)+n(n-2)^2}{b^4}+\frac{2}{a^4}$\\
$\mathcal{R}^{4}_{13}$&$\sim(n+n^2(n-2))\frac{(n-1)^2}{b^4}+\frac{2}{a^4}$
\end{tabular}
\end{center}
These formulae reproduce the table we had before for $n=2$. As before, some contractions
give the same functional dependence on the radii, so it is possible to choose the parameters, such
that one is left with a reduced subset of them. We will not reproduce the attractor equations (\ref{attractor})
for this case, given their length. However, we should say that the behaviour of the system is analogous to
the $n=2$ case.

Corrections involving the fluxes, can be computed in ten dimensions once the action is known. As
it is expected, the attractor equations are formidably lengthy and complicated, so we refrain 
from reproducing them explicitely. 

%%%%%%%%%%%%%%%%%%%%%%%%%%%%%%%%%%%%%%%%%%%%%%%%%%%%%%%%%%%%%%%%%%%%%%%%%%%%%%%%%%%%%%%%%%%%%%%%%%%%%%
\acknowledgments
We are very grateful to A. Sen for his comments on the manuscript. We would also like to thank M. B. Green
and A. Sinha for useful discussions and for reviewing the text.  Finally, we thank A. Castro, Q. Exirifard,  
M. Paulos and P. Vanhove, for their comments. L-Y Hung thanks the Gates Cambridge Trust and L. I. Uruchurtu 
would like to thank CONACyT M\'{e}xico and ORSAS UK for financial support.

\bibliographystyle{JHEP}
\bibliography{numbib}

\end{document}